\documentclass{appolb}
\usepackage{graphicx}
\usepackage{amsmath}
\usepackage{amssymb}
\usepackage{cite}
\usepackage{bbm}
\usepackage[usenames,dvipsnames]{color}
\usepackage[dvipsnames]{xcolor}

\newcommand{\mA}{\mathcal{A}}

\newcommand{\mC}{\mathcal{C}}

\newcommand{\mS}{\mathcal{S}}

\newcommand{\mJ}{\mathcal{J}}

\newcommand{\mT}{\mathcal{T}}
\newcommand{\mK}{\mathcal{K}}
\newcommand{\mN}{\mathcal{N}}
\newcommand{\mO}{\mathcal{O}}
\newcommand{\mW}{\mathcal{W}}

\newcommand{\sct}{\text{\texttt{t}}}

\newcommand{\sctbdy}{\sct^{bdy}(x)}

\newcommand{\ep}{\sigma}

\usepackage{hyperref}

\newcommand{\executeiffilenewer}[3]{%
	\ifnum\pdfstrcmp{\pdffilemoddate{#1}}%
	{\pdffilemoddate{#2}}>0%
	{\immediate\write18{#3}}\fi%
}

\begin{document}
\title{Discrete scale invariance in holography and an argument against the complexity=action proposal.
\thanks{Proceedings for the talk "Discrete scale invarinace in holography revisited" presented at the 6th Conference of the Polish Society on Relativity.}%
}
\author{Mario Flory
\address{Institute of Physics, Jagiellonian University, \\
	\L{}ojasiewicza 11, 30-348 Krak\'ow, Poland}
}
\maketitle
\begin{abstract}
The AdS/CFT correspondence often motivates research on questions in gravitational physics whose relevance might not be immediately clear from a purely GR-perspective, but which are nevertheless interesting. In these proceedings, we summarise two such results recently obtained by the author. One concerns, broadly speaking, the possible isometry-groups of a spacetime sourced by physical matter. The other one provides a possible argument against the recently proposed complexity=action conjecture. 
\end{abstract}
  
\section{Introduction}
\label{sec::Intro}

The AdS/CFT correspondence 
allows to view gravitational physics in terms of a dual quantum field theory interpretation.  This often motivates research on questions in gravitational physics whose relevance might not be immediately clear from a purely general relativity (GR)-perspective, but which are nevertheless interesting. As illustration of this claim, let us look at the complete GR-action in an arbitrary spacetime region $\mW$, following the summary of \cite{Lehner:2016vdi} (see also references therein): 
\begin{align}
\mA\propto&\frac{1}{2}\int_{\mW} \left(R-2\Lambda\right)\sqrt{-g}d^3x 
+\sum_{\mT_i}\int_{\mT_i} K \sqrt{-\gamma}d^2x
+\sum_{\mS_i}\int_{\mS_i} K \sqrt{\gamma}d^2x
\nonumber
\\
&
+\sum_{\mN_i}\int_{\mN_i} \kappa d\lambda \sqrt{\rho}dx
+\sum_{\mJ_i}\int_{\mJ_i} \eta_{\mJ_i} \sqrt{\rho}dx
\nonumber
+\sum_{\mN_i}\int_{\mN_i} \theta \log(|\theta \ell_c|) d\lambda \sqrt{\rho}dx
\nonumber
\end{align} 
Herein, only the first term dates back to Einstein and Hilbert, 
 the second and third are the necessary boundary terms on spacelike and timelike boundaries $\mT_i,\mS_i$ of $\mW$ derived in the 70'ies, while the remaining terms were studied in more modern times, and describe the roles of joints $\mJ_i$, null boundaries $\mN_i$, and so-called counter terms (the last term) giving the action $\mA$ a well-defined reparametrization-invariant value \cite{Lehner:2016vdi}. Hence, despite the centennial of General Relativity, a complete picture of the gravitational action $\mA$ was only recently formulated, and this work was strongly motivated by AdS/CFT, specifically the recent complexity=action (CA) conjecture \cite{Brown:2015bva}, to which we will return in section \ref{sec::CA}.

\section{Discrete scale invariance in holography}
Consider the question whether for a given (Lie)group there exists a smooth (semi)-Riemannian metric with that group as its isometry group, which in Einsteins equations can be sourced by matter satisfying the weak energy condition (and possibly also satisfies additional requirements). This seems like a question, if maybe a bit obscure, belonging squarely into the realm of GR and differential geometry. But in holography it takes on a life of its own: A well-known corner-stone of the AdS/CFT correspondence is that the isometry group of $AdS_{d+1}$ corresponds to the conformal group of a $CFT_d$. As argued in \cite{Balasubramanian:2013ux}, breaking the isometry group of AdS down to the Poincar\'{e}-group combined with a \textit{discrete scale invariance (DSI)} would correspond to the holographic description of a cyclic RG-flow - a highly unusual and potentially very interesting phenomenon. In fact, two models seeming to do just that were presented in \cite{Balasubramanian:2013ux}, one "bottom-up" and one "top-down". These models gave rise to solutions to Einsteins equations of the form 
\begin{align}
ds^2=e^{2C(w,\theta)}\left(e^{2w/L}\left(-dt^2+d\vec{x}^2\right)+dw^2\right)+e^{2B(w,\theta)}\left(d\theta+A(w,\theta)dw\right)^2\textcolor{white}{\frac{1}{2}}
\nonumber
\end{align}
with specific periodic functions $A,B$ and $C$, such that only Poincar\'{e}-invariance in the boundary directions $t,\vec{x}$ combined with DSI is manifestly preserved.   

However, as pointed out in \cite{Flory:2017mal}, the only way to \textit{prove} that the AdS-isometry group was successfully broken down to a subgroup is to analytically find all linearly independent solutions of the Killing-equations 
\begin{align}
\nabla_{\mu}\mK_{\nu}+\nabla_{\nu}\mK_{\mu}=0
\nonumber
\end{align}
in the metric above, and checking the isometry algebra formed by the Lie-brackets of these Killing vector fields. 
This was done in \cite{Flory:2017mal}, proving that the bottom-up model of \cite{Balasubramanian:2013ux} still exhibits a full $AdS_d$-like isometry group, and hence does not describe genuine DSI. However, the top-down model of \cite{Balasubramanian:2013ux} does seem to describe genuine DSI, and hence deserves further study.

\section{An argument against the complexity=action proposal}
\label{sec::CA}

Let us now return to the Einstein action written down in section \ref{sec::Intro}, specifically the counter term. As pointed out in \cite{Flory:2019kah} using Raychaudhuri's equation, in a $2+1$-dimensional vacuum spacetime, this term is a total derivative and can be rewritten as 
\begin{align}
\mA_{counter}&=\iint_{\lambda_{min}}^{\lambda_{max}} \left(\partial_\lambda \sqrt{\rho}\right) \log(|\theta \ell_c|) d\lambda dx=\int\left[\sqrt{\rho}\log(|\theta \ell'_c|) \right]\Big|_{\lambda_{min}}^{\lambda_{max}}dx
\label{counter}
\end{align}
where $\lambda$ is a coordinate along the null-rays foliating the null-boundary $\mN_i$, $x$ is a coordinate enumerating these null-rays, $\sqrt{\rho}$ is a volume-element along spacelike curves parametrized by $x$, and the expansion is defined as $\theta=\frac{1}{\sqrt{\rho}}\partial_\lambda \sqrt{\rho}$. Thus, the contribution from this term can be written as an integral along the joint-curves where null-boundaries of $\mW$ collide with other boundaries, or each other. This has an important consequence \cite{Flory:2019kah}: When applying an infinitesimal \textit{local conformal transformation} (with infinitesimal expansion parameter $\sigma\ll1$) to the simple Wheeler-De Witt patch in $AdS_3$ which corresponds to the vacuum state of the dual theory, the leading behaviour of the change of the action will come from this term and read 
\begin{align}
\mA_{counter}\sim\sigma\log{\sigma}.
\label{slogs}
\end{align}

\begin{figure}[htb]
\begin{center}                                      
	\def\svgwidth{0.99\columnwidth}
	\executeiffilenewer{tent.svg}{tent.pdf}%
	{inkscape -z -D --file=tent.svg %
		--export-pdf=tent.pdf --export-latex}%
	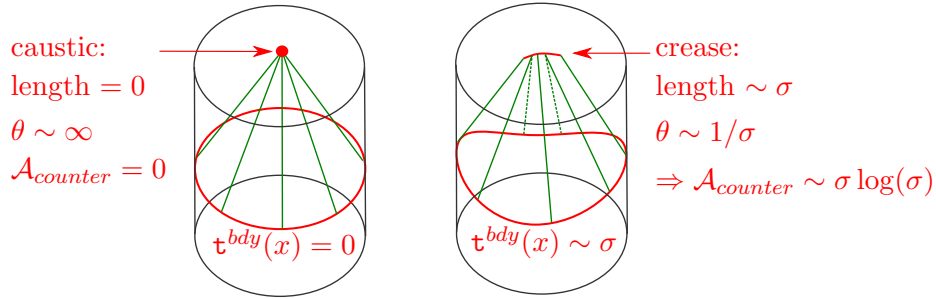%

\end{center}
\caption{The cylinders represent global $AdS_3$, with time coordinate increasing upwards. As explained in \cite{Flory:2019kah}, the infinitesimal local conformal transformation on the boundary holographically corresponds to a large diffeomorphism in the bulk, which can be understood to deform a boundary equal-time-slice (red) from $\sctbdy=0$ (left) by a fluctuation of order $\sigma$ (right). This naturally breaks translation invariance and leads to  a crease in the null-boundary of the WdW patch $\mW$ (lightrays of the future null-boundary are shown in green) of length of order $\sigma$, along which the integral \eqref{counter} has to be carried out. The expansion $\theta$ at this crease is of order $1/\sigma$, as in the limit $\sigma\rightarrow 0$ the crease becomes a caustic, at which $\theta\rightarrow\infty$.}
\label{fig::tent}
\end{figure}

 We will not repeat the precise calculations here that lead to this result in \cite{Flory:2019kah}, but instead we will give a qualitative argument for this result in figure \ref{fig::tent}. 

This result has very immediate consequences for the CA-proposal of \cite{Brown:2015bva}, which claims that the value of the gravity-action $\mA$, evaluated on a Wheeler-De Witt patch $\mW$, is holographically equivalent to a measure of complexity $\mC(\psi)$ of the dual field theory state $\psi$. Complexity herein is understood as a distance measure on the space of states, and hence is subject to consistency requirements such as positivity, the triangle inequality, etc.~\cite{2005quant.ph..2070N}
\footnote{Technically, as in \cite{Flory:2019kah} we assume that complexity is primarily defined as a distance measure on the space of unitariy operators, which after a choice of reference state induces a distance measure on the space of states that can be reached from the reference state by unitary transformations.}.

However, we can now show a contradiction between the following three assumptions (for $\ep\ll1$): 
\begin{enumerate}
	\item The infinitesimal local conformal transformation is generated by a unitary operator $U(\ep)=\mathbbm{1}+\ep V +\mO(\ep^2)$ ($V$ can be explicitly expressed in terms of Virasoro generators) with complexity $\mC(U(\ep))=\ep \mK'+\mO(\ep^2)$, $0<\mK'<+\infty$. The latter condition follows from the positive homogeneity property of Nielsen's proposal \cite{2005quant.ph..2070N} for complexity with the additional assumption of using  finite "penalty factors".
	
	\item The change of complexity of the dual state $\psi$ caused by applying the operator $U$ has to be less than the complexity of $U$: $\mC(U(\ep))\geq|\delta\mC(\psi)|$ (Triangle inequality).
	
	\item $|\delta\mC(\psi)|=\mK |\sigma\log(\sigma)|,\ 0<\mK<+\infty,$ (equation \eqref{slogs}) because of the CA proposal when using the action as given in section \ref{sec::Intro}.
\end{enumerate}
Together, these three assumptions would imply that $\ep \mK'\geq \mK |\sigma\log(\sigma)|$ as $\ep\rightarrow0$\ for positive finite constants $\mK$ and $\mK'$, which is false. Hence, any definition of field-theory complexity that satisfies assumptions 1 and 2 can not be exactly dual to the CA proposal in AdS$_3$/CFT$_2$ with the counter-terms chosen as in section \ref{sec::Intro} (which implies assumption 3 by equation \eqref{slogs} as shown in \cite{Flory:2019kah}). 

In fact, a similar argument can be applied to the complexity-change caused by time-evolution by an infinitesimal time-step $\delta t$ beyond the critical time $t_c$ in \cite{Carmi:2017jqz}. This would lead similarly to a complexity change of the form $|\delta\mC(\psi)|=\mK |\delta t\log(\delta t)|$, and the same contradiction as above could be constructed. Interestingly, in this case the contradiction even arises in general dimensions and independently of whether the counter term is added to the gravity action or not, however it was also argued in \cite{Carmi:2017jqz} that this problem can be solved by requiring the complexity as a function of time to be smoothed out over a certain time scale.

\section*{Acknowledgements}

This research was supported by the Polish National Science Centre (NCN) grant 2017/24/C/ST2/00469.

\bibliography{main}{}
\bibliographystyle{utphys}

\end{document}